\documentclass{iau}
\usepackage{graphicx}

\title[100~GHz Receiver Spectrometer] 
{100 GHz Room-Temperature Laboratory Emission Spectrometer}

\author[Nadine Wehres et al.]   
{Nadine Wehres,
Bettina Heyne,
Frank Lewen,
Marius Hermanns,
Bernhard Schmidt,
Christian Endres \thanks{Present address: Max-Planck Institut f\"ur extraterrestrische Physik, Giessenbachstrasse 1, 85748 Garching, Germany},
Urs U. Graf,
Daniel R. Higgins,
\and Stephan Schlemmer
}

\affiliation{I. Physics Institute, University of Cologne, \\
Z\"ulpicher Str. 77, 50937 Cologne, Germany \\ email: {wehres@ph1.uni-koeln.de; schlemmer@ph1.uni-koeln.de} \\[\affilskip]
}

\pubyear{2017}
\volume{332}  
\jname{Astrochemistry VII, Through the Cosmos from Galaxies to Planets}
\editors{M. Cunnningham, T. Millar \& Y. Aikawa, eds.}
\begin{document}

\maketitle

\begin{abstract}
We present first results of a new heterodyne spectrometer dedicated to high-resolution spectroscopy of molecules of astrophysical importance.  The spectrometer, based on a room-temperature heterodyne receiver, is sensitive to frequencies between 75 and 110 GHz with an instantaneous bandwidth of currently 2.5 GHz in a single sideband.  The system performance, in particular the sensitivity and stability, is evaluated. Proof of concept of this spectrometer is demonstrated by recording the emission spectrum of methyl cyanide, CH$_3$CN. 
Compared to state-of-the-art radio telescope receivers the instrument is less sensitive by about one order of magnitude.  Nevertheless, the capability for absolute intensity measurements can be exploited in various experiments, in particular for the interpretation of the  ever richer spectra in the ALMA era. The ease of
operation at room-temperature allows for long time integration, the fast response time for
integration in chirped pulse instruments or for recording time dependent signals.  Future prospects as well as limitations of the receiver for the spectroscopy of complex organic molecules (COMs) are discussed.
\end{abstract}

\section{Introduction}

 High-resolution rotational spectroscopy in the laboratory enables us to obtain intrinsic molecular information, such as rotational constants, as well as (higher order) distortion constants with which precise transition frequencies for molecules can be predicted.  This knowledge allows astronomers to model spectra of molecules under different physical conditions, i.e., different temperatures, which in turn enables identification in the observational data sets.  This work becomes more and more demanding as the data sets include thousands of lines of many different species. The role of istopologes, vibrational excited states, isomers and even molecular conformers, but also the role of internal molecular excitation, are subjects of current research.
 
 The field of molecular astrophysics strengthens our knowledge on the constituents of the interstellar and circumstellar medium (ISM/CSM) and deepens our understanding of the physical conditions in, for example, star forming regions.  With the identification of molecules in the ISM/CSM we obtain information on temperatures, column densities, and abundances of such species.  So called unbiased spectral line surveys can furthermore give insight into the details of chemical reaction networks, especially on the formation and destruction pathways of molecules within a certain astronomical environment.  Molecules are ubiquitous in the ISM/CSM and the field of astrochemistry is closely linked to the progresses in the field of molecular physics in the laboratory and observational astronomy.

Millimetre and sub-millimetre (mm/submm) telescopes use very sophisticated receiver technology for the detection of the weak rotational emission spectra.  The most recent developments in observational astronomy are large space based and ground based telescopes operating at frequencies of 80~GHz up and well into the THz regime, e.g., ALMA (the Atacama Large Millimeter/submillimeter Array), APEX (the Atacama PAthfinder Experiment), as well as SOFIA (the Stratospheric Observatory for Far-Infrared Astronomy), and the JWST (James Webb Space Telescope) that soon will be launched in October 2018.  Observational Astronomy has therefore pushed forward the technological advances, leading to more sensitive receivers with higher spectral (and spatial) resolution covering up to 16~GHz bandwidth at present (\cite{Klein:2012}).
In this study we take advantage of the most recent developments in receiver technology that mainly has been pushed forward by the astronomy community.  Room-temperature low-noise amplifiers operating at up to 300~GHz finally enable direct amplification of the weak molecular signal without lossy down-conversion of the high frequency signal first.  Hence, the new advances in mm/submm technology become competitive with traditional frequency modulation (FM) absorption spectroscopy in the laboratory.

Here, we present the very first room-temperature heterodyne receiver dedicated to high-resolution gas phase spectroscopy of complex organic molecules in the laboratory.
The advantage of using the new heterodyne spectrometer lies in the fact that receivers enable emission spectroscopy instead of using the traditional way of scanning frequency absorption spectroscopy.  Heterodyne receiver detectors for the detection of weak emission signals of complex molecules offer a better understanding of intrinsic physical parameters such as lineshapes and absolute transition intensities.  Such information can hardly be obtained by using the traditional way of FM absorption spectroscopy.
Moreover, faster detection schemes are enabled by using emission spectroscopy, since spectra over several GHz of bandwidth can be detected instantly by using a sensitive receiver coupled to Fast Fourier Transform Spectrometers as backends.

Bringing these new technological advances to the laboratory will have the following advantages over traditional absorption technology using Schottky diodes or bolometers as detectors:
\begin{itemize}
\item absolute intensity calibration
\item intensities reflecting LTE (local thermal equilibrium) conditions
\item large instantaneous bandwidth of several GHz in one shot
\item proper lineshapes/no derivatives
\item faster detection schemes
\item possible single sideband operation
\item extension to quantum limited detectors is possible
\end{itemize}

All these advantages are now permitted by using radio astronomical receivers.  Different experiments have been used in the past to also obtain broad instantaneous frequency coverage providing high spectral resolution.  Especially test laboratories for the SWAS (Submillimeter Wave Astronomy Satellite), Odin and the Herschel satellite used a similar approach (\cite{Tolls:2004,Teyssier:2004,Frisk:2003,Higgins:2010}).  However, all these experiments were solely used to understand and test the experimental set-ups and to obtain a pre-launch calibration for the onboard receivers.  Due to the large complexity of the instruments none of these experiments used astronomical receivers with the sole purpose of focussing on laboratory studies.  However, \cite{Tanarro:2017} used the CNTRAG-IGN telescope (Guandalajara, Spain) in combination with a laboratory glass-cell to detect molecular emission lines for laboratory studies at frequencies between 41--49~GHz. Also other groups in Japan, Spain, the UK and the USA are pushing for the advancement of laboratory emission spectrometers. 

In the following, first the instrument will be described, then the performance will be tested and checked against its design figures such as noise temperature, stability and sensitivity. In the Results section first spectra will be shown and compared to predictions based on previous knowledge. Finally, the laboratory emission spectrometer will be evaluated and future prospects will be discussed.

\section{Experiment}

\begin{figure}
\begin{center}
\includegraphics[width=1\textwidth]{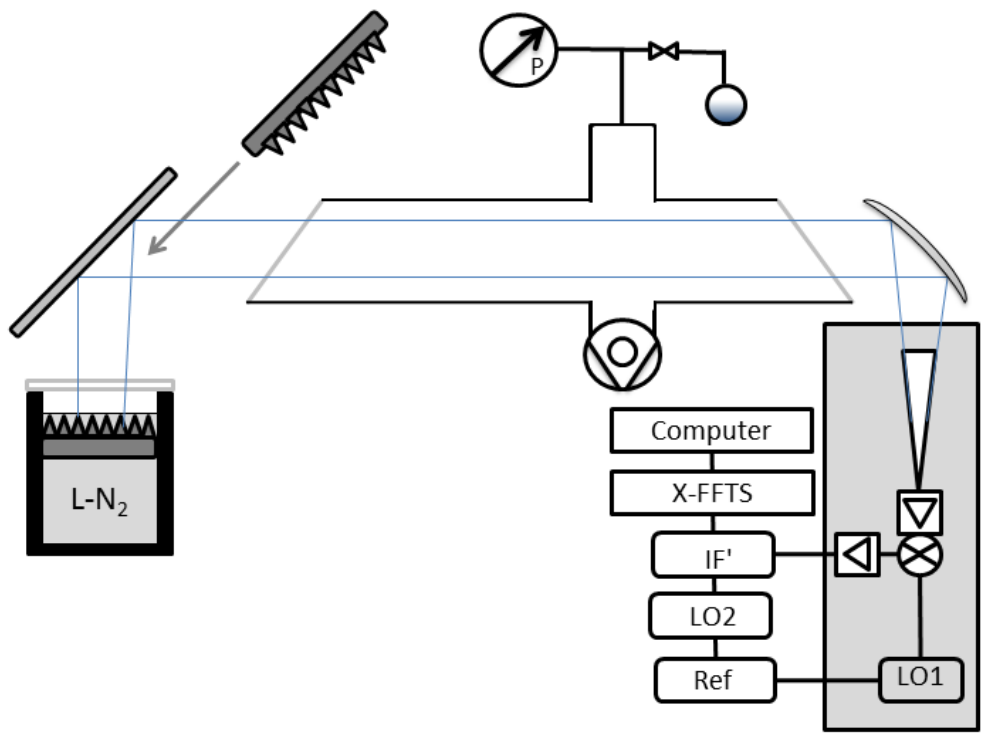}%
\caption{\label{fig:spectrometer} Schematic diagram of the 100~GHz room-temperature laboratory emission spectrometer.  Light in the beam profile of the spectrometer (blue lines) is accepted by the receiver antenna. Light from molecules in the vacuum chamber is recorded against the liquid nitrogen continuum signal (cold load).
The cold load consists of absorber foam immersed in liquid nitrogen acting as a
blackbody emitter. For absolute intensity calibration a room-temperature absorber
foam (hot load) is placed in between the chamber and the second mirror.  A dual pumping stage allows for constant pressure of around 1$\times$10$^{-2}$~mbar during the experiment.  The molecular
sample, here methyl cyanide (CH$_3$CN), is located in a glass flask and admitted to the vacuum chamber through a needle valve. The grey shaded area underneath the receiver indicates cooling by a Peltier element of the copper block receiver mount installed for thermal stability of the sensitive detector.  More details on the receiver electronics and the backend detector are given in the text.  }
\end{center}
\end{figure}

\subsection{Laboratory Emission Spectrometer Set-Up}

The complete laboratory emission spectrometer set-up is shown in Fig.~\ref{fig:spectrometer}. 
 The receiver is the heart of the instrument and shown on the right side of Fig.~\ref{fig:spectrometer} and will be described in more detail in the following subsections.
 It detects light from the emitting molecules located in a pyrex vacuum cell against the cold background depicted in the left part of Fig.~\ref{fig:spectrometer}. The light is directed using a flat mirror (left part) and an elliptical mirror (right part). 
 The cold load consists of a Dewar with a cold absorber foam at liquid nitrogen temperatures ($\simeq$ 77~K).
 For calibration purposes a hot load (room-temperature absorber foam) can be placed in between the cold load and glass cell (just in front of the flat mirror), as shown in Fig.~\ref{fig:spectrometer}.
 
 A pressure gauge (Leybold, CERAVAC) is used to control the pressure inside the chamber upon gas injection.  The molecules are located in a glass flask.  Here, methyl cyanide (CH$_3$CN) is used.  A needle valve allows for precise flow of the sample molecules into the cell.  The total length of the glass chamber is around 2~m.  On each side of the glass-cell HDPE (high density polyethylene) windows are mounted at Brewster angle.  The diameter of the glass cell is about 10~cm.  During the experiment the pressure is held constant.  Typical pressures are around 1~x~10$^{-2}$~mbar.  For this, a dual pumping stage, consisting of a rotary vane pump (Leybold-Heraeus, 8~m$^3$ hr$^{-1}$) and a turbomolecular pump (Leybold-Heraeus 150~l~s$^{-1}$) is used.  After each data acquisition step, which usually lasts about 10 minutes, the glass chamber is pumped out and refilled again with $"$fresh$"$ molecules.
As will be discussed in the Results section temperature stability of the receiver is an issue of concern. Therefore we implemented a Peltier cooling element just underneath the receiver.  This is indicated by the grey shaded area in Fig.~\ref{fig:spectrometer}.  The Peltier element keeps the receiver at a constant temperature and prevents warming-up of the electronics during operation. The Peltier is usually kept at temperatures around 15$^\circ$C.

\begin{figure}
\begin{center}
\includegraphics[width=0.5\textwidth]{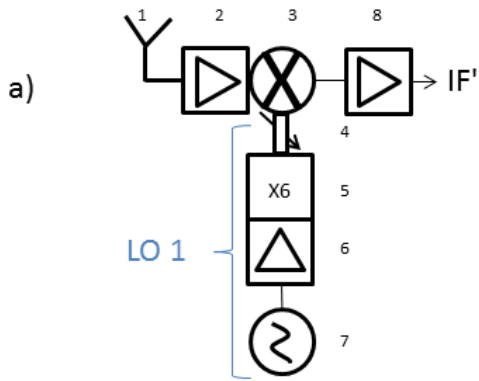}
\caption{\label{receiver} Electronics scheme of the room-temperature heterodyne receiver.  The receiver consists of the corrugated horn antenna (1), followed by a W LNA (2), and a mixer (3), operational between 75-110~GHz.  The mixer downconverts the molecular signal using the LO~1 signal (7). The LO~1 signal is based on a synthesizer signal which is amplified (6) and multiplied (x6) (5).  To match the conversion loss a waveguide tuneable attenuator is used (4).  The mixer output is amplified using an LNA (Miteq, +34~dB) (8) before entering the IF processor.  }
\end{center}
\end{figure}

\subsection{Frontend:  The Heterodyne Receiver}
The new laboratory receiver depicted in Fig.~\ref{fig:spectrometer} is shown in more detail in Fig.~\ref{receiver}. The signal is received by a corrugated horn antenna (WR10, RPG company, 1) connected to a state-of-the-art low noise amplifier (W LNA 75--110, RPG company, 2) with a gain of 20 dB and a noise figure of 4~dB. The resulting signal is then fed into a fundamental balanced mixer (FBM, RPG company, 3) together with the signal from a local oscillator (7).  The conversion loss of the mixer is 9.5~dB.  The mixer multiplies the high frequency signal of the molecules (around 75--110 GHz) with the LO signal of a synthesizer which is synchronized to a 10~MHz Rubidium clock.  The synthesizer is set to an output power of 6.5~dBm, and its output is guided into an amplifier and multiplier chain (AFM6, RPG company, 6 and 5) resulting in about +10~dBm output power.  A waveguide tuneable attenuator (WTA-110, RPG company, 4) with an attenuation range between 0.4 and 40~ dB is used to match the conversion loss in the mixer.  The intermediate frequency signal is then further amplified by a room-temperature low noise amplifier (LNA, Miteq company, +34 dB, 8) before entering the IF-processor as a signal indicated as IF$^\prime$.
The few components consisting of synthesizer (7), amplifier-multiplier chain (6, 5), as well as tuneable attenuator (4) are combined and called local oscillator~1 (LO~1) for future reference in the text.

\begin{figure*}
\begin{center}
\includegraphics[width=1\textwidth]{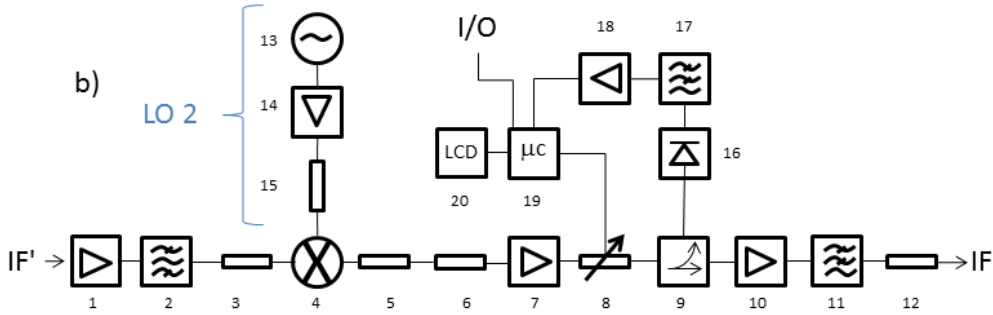}
\caption{\label{IF} The intermediate Frequency Processor.  The intermediate frequency signal (IF$^\prime$) is amplified (25~dB, 1) and a bandpass filter (2) allows for frequencies of 6.35 $\pm 1.25$~GHz to pass.  The signal is attenuated by 7~dB (3) before reaching the second mixing stage (4).  Here, the IF$^\prime$ is multiplied with the output of a 7.6~GHz direct digital syntheizser (DDS, 13), synchronized to a Rubidum clock.  The LO (13) is then amplified by 24~dB (14) and attenuated by 3~dB (15).  The mixing process results in frequencies of 7.6~GHz $\pm$ 6.35~GHz (4).  The IF then gets attenuated (2 x 3~dB, 5 and 6), and amplified by +24~dB (7).  A digital step attenuator (maximum 31~dB, 8) is controlled by a microcontroller (19).  A small portion (about 1\%) of the signal is coupled out by a directional coupler (9).  Here, we obtain a continuum signal over the whole frequency range by using a continuum detector (tunnel diode, 16).  A lowpass filter is applied for better signal stability (17).  The counts are amplified (18) and a micro-controller ($\mu$C, 19) outputs the signal onto an LC-display (20).  The remaining signal (about 99 \%) passes the directional coupler (9), gets amplified (24~dB, 10), is send through a lowpass filter (loss 1~dB, 11), gets attenuated (7~dB, 12), and is fed into the XFFTS.}
\end{center}
\end{figure*}

\subsection{Backend:  IF-Processor and XFFTS}
The intermediate frequency (IF$^\prime$) is further processed by a home made IF-processor to match the input requirements of the XFFTS (eXtended Fast Fourier Transform Spectrometer) (\cite{Klein:2012,Klein:2006}).  A schematic overview of the IF-processor is shown in Fig.~\ref{IF}.  Here, the IF$^\prime$ signal is once more amplified using a Miteq amplifier (25~dB, 1), and filtered through a bandpass filter (Reactel, Inc., loss -1 dB, 2).  This will give signals at IF$^{\prime}$ of 6.35 $\pm$~1.25~GHz.  We employ a 7~dB attenuator (3) before the signal enters a second heterodyne mixing stage (Marki, conversion loss -6.5~dB, 4).  Here, the IF$^\prime$ signal is multiplied with an LO frequency of 7.6~GHz (5~mW, 13), also synchronized to a 10~MHz Rubidium clock.  The LO (13) is amplified (24~dB, 14), and attenuated (3~dB, 15), before entering the second mixer stage (4).  This results in a center frequency of 1.25~GHz with a bandwidth of $\pm$~1.25~GHz.  The resulting frequency (IF) is then attenuated (2 x 3~dB, 5 and 6), and amplified (Mini Circuits, 24~dB, 7).  A digital step attenuator (Mini Circuits, up to 31~dB, 8) mounted in front of a directional coupler (Mini Circuits, 9) matches the power level to the input requirements of the XFFT-spectrometer.  The directional coupler takes about 1$\%$ off of the IF signal for continuum detection (16), the remaining $\sim~99\%$ are amplified (Mini Circuits, 24~dB, 10), guided through a 2.5~GHz lowpass filter (-1~dB, 11), attenuated by 7~dB (12) and the resulting signal at 0.1--2.5~GHz is then detected by the XFFTS.  The continuum signal that is coupled out in step (9) is detected by a continuum diode detector (16) with internal attenuation of 6~dB.  An RC-lowpass filter (17) is applied for pre-integration, and an amplifier (18) is used before the signal is processed by a micro-controller (19).  The output is shown on an LC-display (20), which also shows the applied digital attenuation (8), as well as the continuum signal.  In total the molecular signal gets thus amplified by a factor of roughly 100~dB.
Overall, we use two mixers.  As described before, the first mixer is pumped by the so-called LO~1 stage included in the receiver itself (see Fig.~\ref{receiver}).  The second mixer is included in the IF-processor and pumped by the LO~2, which is labelled in Fig.~\ref{IF}, and includes parts 13, 14, and 15.

\section{System Performance}
The receiver described in the preceeding section is operated very similar to the receivers of a radio telescope, except the molecules are held in the vacuum cell. Our receiver is held at room-temperature while in contrast the ALMA band 3 receivers are held at some 4~K with a system noise temperature around 40~K. Therefore it is interesting how the warm detector might compete with the complex cryogenic temperature detector. For that reason the noise temperature as well as the system stability of the laboratory emission spectrometer will be tested in the following.

\subsection{Noise Temperature}
The receiver and system noise temperatures were evaluated using the so-called Y-factor method.

\begin{equation}
\rm{T_{noise}=\frac{T_{hot}-Y~T_{{cold}}}{Y-1}}
\end{equation}

Y denotes the ratio of the power measured when detecting the hot and cold loads, respectively:

\begin{equation}
\rm{Y=\frac{P_{hot}}{P_{cold}}~\simeq~\frac{T_{hot}}{T_{cold}}}
\end{equation}

\begin{figure}
\begin{center}
\includegraphics[width=\textwidth]{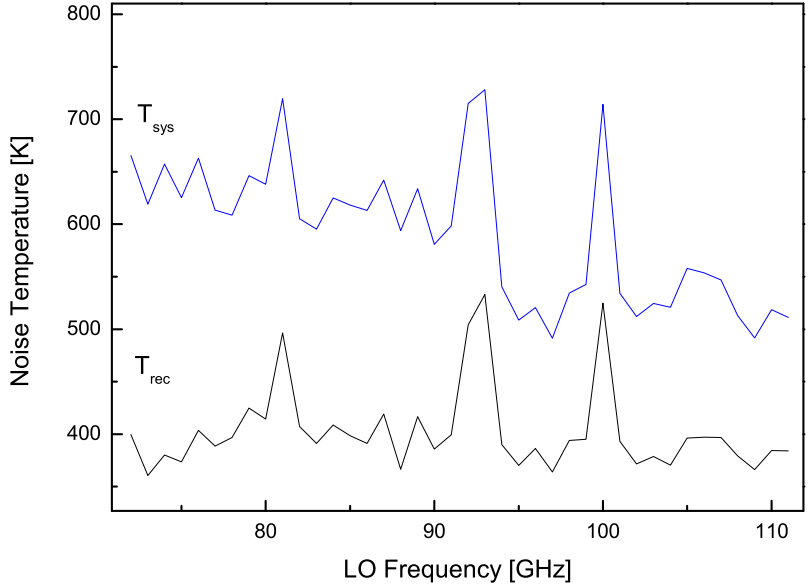}
\caption{\label{fig:noise} The receiver and system noise temperatures have been evaluated using the Y-factor method applied to the continuum signal detector.  The figure shows the frequency dependence of the noise temperature of the receiver, T$_{rec}$, only (black curve), as well as the receiver plus optical path and vacuum chamber, T$_{sys}$ (blue curve). }
\end{center}
\end{figure}

For absolute intensity calibration we use microwave absorber foam at room temperature (293-300~K) as well as at 77~K (liquid nitrogen temperature), respectively.  For obtaining the receiver noise temperature the absorber is placed in front of the corrugated horn antenna, and the continuum output signal is recorded as a function of frequency.  The result is shown in Fig.~\ref{fig:noise}.  The LO~1 frequency is varied over the bandwidth of the detector between 72 and 112~GHz, in steps of 1~GHz with input power levels of 6.5~dB in order to optimize the conversion gain.  The black curve shows the receiver noise temperature (T$_{rec}$) as determined by the continuum diode detector described above when the absorber is placed in front of the antenna.

The blue curve shows the system noise temperature (T$_{sys}$) when the absorber is placed at the end of the glass cell (see also Fig.~\ref{fig:spectrometer}).  In total, T$_{sys}$ varies between 650 and 550~K over the whole frequency range (72 -- 112~GHz).  Here, three temperature spikes are visible at frequencies of 81, 91-92, and 101~GHz.  Assumingly, the temperature spikes arise from intermodulation effects caused by non-linearities in the system.
Furthermore, we see a clear slope in T$_{sys}$ (blue curve, which includes the vacuum chamber).  The slope is reproducible and most likely due to a non-optimal focussing mirror allowing the beam to spread too far and hit the glass-cell walls.  This vignetting will then cause a slope in the detected noise temperature with an increase towards lower frequencies, as can be seen in the blue curve.

T$_{rec}$ is significantly lower than T$_{sys}$ and yields temperatures in the order of 400~K (black curve).  Here, the same temperature spikes are detected independent of whether the glass-cell is used or not. T$_{sys}$ (including vacuum chamber, HDPE windows, and mirrors, blue curve) is roughly 200~K higher.  We determined that about 40~K are due to emission of the HDPE windows of which each contributes about 20~K to the total noise temperature.

For comparison, the theoretical value for the noise temperature of the receiver is calculated by taking into account the different noise figures given on the data sheets of the respective electronic components.  According to the specifications, the  main noise contribution is due to the low noise amplifier (LNA) with a noise figure of NF~=~4~dB.  The gain of the LNA is G~=~20~dB.  Hence, for the LNA by itself a noise temperature of 438~K, is expected, according to:

\begin{equation}
\rm{T}_{LNA}=290~K \cdot (10^{\frac{\rm{NF}}{10}} -1)
\end{equation}

The receiver noise temperature can then be calculated to be:

\begin{equation}
\rm{T}_{rec}=\rm{T}_{LNA}+ \frac{\rm{T}_{Mixer}}{\rm{G}_{LNA}}
\end{equation}
\begin{equation}
\rm{T}_{rec}=461~K
\end{equation}

The mixer losses are not included here, but are in the order of 9.5~dB and will likely increase T$_{rec}$ even further.  However, according to our measurements shown in Fig.~\ref{fig:noise} the receiver seems to perform slightly better than expected by the specifications.
This result shows that the receiver noise temperature is about a factor of 10 worse than e.g. an ALMA band 3 receiver. However, system operation at room temperature is much easier and allows even for student education as it was the case here.

\subsection{System Stability}
\begin{figure}
\begin{center}
\includegraphics[width=\textwidth]{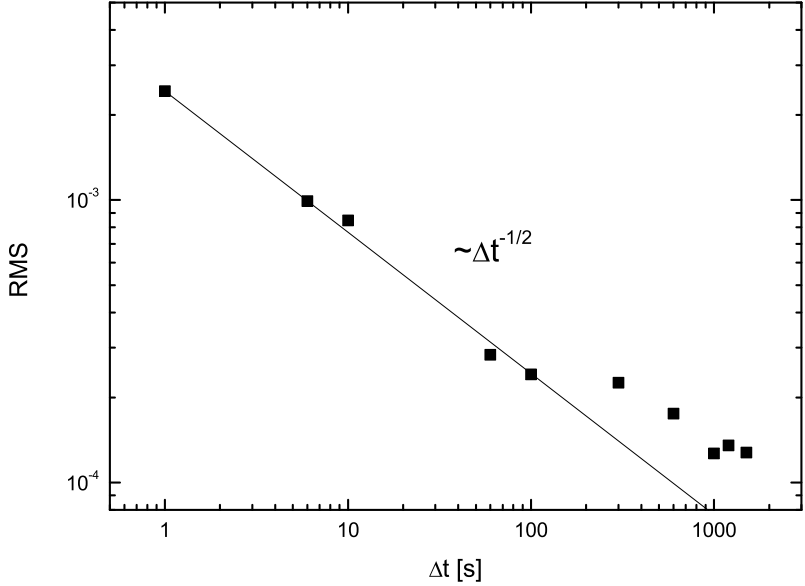}
\caption{\label{fig:stability} The stability of the spectrometer was evaluated from the background signal while taking spectra of methyl cyanide. The RMS value of the background signal was determined with respect to the integration time and a fixed channel width of 76~kHz (no binning).  The integration times ($\Delta$t) were varied between 1 and 1500~s. The black line shows the expected improvement for statistical noise. }
\end{center}
\end{figure}

In order to obtain information on the system stability (so-called Allan stability) we recorded emission spectra of methyl cyanide for different integration times and fixed channel width of 76~kHz (no binning, about 88~kHz spectral resolution).  In Fig.~\ref{fig:stability} the standard deviation of the baseline noise (RMS value) is plotted as function of the integration time.  The integration time was varied between 1 and 1500 seconds.  For integration times up to 100~s the improvement of the RMS with respect to the integration time is as expected and follows an inverse square-root dependence.  The black line shows this expected improvement for statistical white noise.  For integration times larger than 100~s the improvement deviates from the expected behavior and gets worse.  Assumingly, this is due to drifts in the electronics during the experiments. This is not unexpected as the net amplification amounts to some 100~dB and is the result of many amplification and attenuation stages. Drifts occur naturally when the experimental set-up slowly warms up throughout the experiments.  This causes variations in the continuum signal level.  For this reason the receiver is kept at a constant temperature using a Peltier element.  However, the intermediate frequency processor is not stabilized in our experiment.
In order to avoid large temperature fluctuations throughout the measurements we kept integration times short (around 600 seconds at most), left the IF-processor constantly switched on, and stabilized the temperature in the laboratory (air conditioning).  For better system stability it might be worthwhile to, e.g., introduce a temperature controlled water re-circulator to the IF processor.  This could potentially result in longer Allan times.

\subsection{Spectrometer Sensitivity}
Once the system noise temperature and Allan stability are determined, we can calculate the sensitivity of the spectrometer by applying the radiometer formula:

\begin{equation}
\rm{\Delta T=\frac{T_{sys}}{\sqrt{\Delta \nu \Delta t}} \times 2}
\end{equation}

Here, T$_{sys}$ is the system noise temperature, $\Delta$t is the integration time, and $\Delta \nu$ is the channel width which is given by the XFFTS.
The factor 2 arises from the fact that the receiver is operated in double sideband mode where the RF signal from both sidebands are combined for each IF frequency. Here we assume a 1:1 sideband ratio.
Using a T$_{sys}$ of around 600~K as a conservative value (see Fig.~\ref{fig:stability} with LO frequency of 86~GHz), and a 100 minute (6,000~s) integration time, as well as a channel width of 76~kHz (channel spacing of the XFFTS at a bandwidth of 2.5GHz), the resulting RMS noise value amounts to about 60~mK.  Successive spectra have been obtained by integrating 10 spectra for about 600~s each.  The spectra where then co-added to arrive at the noise equivalent temperature. This system noise equivalent temperature can be compared to the noise in calibrated spectra as will be described below.

In general, for each emission spectrum of 600~s duration, two separate spectra have to be recorded, i.e., the spectrum of the warm molecular gas (at room-temperature) in front of a cold background (liquid N$_2$), as well as a hot background reference spectrum for absolute intensity calibration, i.e., we record spectra at 77~K and 293~-~300~K, respectively.
The signal S$_{hot}$ is proportional to the power P$_{hot}$ impinging on the receiver which is proportional to T$_{hot}$ in case of the hot (293-300~K) background plus the signal from the receiver noise, i.e., proportional to T$_{sys}$. Thus,
 \begin{equation}
\rm{S}_{hot} ~\sim~\rm{T}_{hot} + \rm{T}_{sys} 
\end{equation}
holds. Similarly, the signal S$_{cold}$  for the cold (77~K) background measurement (without molecules) becomes
 \begin{equation}
\rm{S}_{cold} ~\sim~\rm{T}_{cold} + \rm{T}_{sys}.
\end{equation}

In fact from such measurements the system noise temperature for each IF frequency channel can be derived by solving the above equations for T$_{sys}$. When repeating these measurements (hot and cold) but with molecules in the cell the temperature calibrated signal becomes
 
 \begin{equation}
 \label{signal_analysis}
\frac{\rm{S}_{cold} - \rm{S}_{hot}}{\rm{S}_{hot}}~\times~(\rm{T}_{hot} + \rm{T}_{sys}) .
\end{equation}

Due to construction of this formula the positive molecular signal emerges on top of a negative background of size -(T$_{hot}$ - T$_{cold}$). 
In these first measurements the signal was also subject to some baseline modulation issues based on standing waves. Therefore low frequency signals were corrected by twofold Fourier Transform with rejection of the very low frequency part including the constant background. As a result the negative background was rejected and the molecular signal was unaffected and could be further analysed based on Eq.~\ref{signal_analysis}. Consequent results are discussed in the following section. 

A more thorough investigation of the signal shows that an analysis according to 
\begin{equation}
\frac{\rm{S'}_{cold} - \rm{S}_{cold}}{\rm{S}_{hot} - \rm{S}_{cold}  } ~\times~(\rm{T}_{hot} - \rm{T}_{cold})
\end{equation}

where the signals S'$_{cold}$ and S$_{cold}$ refer to the signal with and without molecules in the cell are less prone to fluctuations and baseline effects. However, three independent measurements are needed for a complete molecular spectrum.

\section{Results and Discussion}

As a proof of concept for the laboratory emission instrument spectra of methyl cyanide, CH$_3$CN, have been recorded. The molecule is per definitionem a so-called complex organic molecule.  Following the convention of \cite{Herbst:2009} complex organic molecules are defined as a molecule containing 6 or more atoms including carbon and hydrogen . Rotational lines of many of those molecules dominate the rich spectra taken by ALMA and other telescopes.  The dipole moment of  CH$_3$CN is quite strong with about 3.9~Debye and makes this molecule a perfect test candidate for our new emission spectrometer.   In terms of rotational spectroscopy this molecule is a symmetric top molecule, in which the rotational constant is about 9.2~GHz, giving rise to spacings of the major rotational features by 2~$B$~$\sim$~18.4~GHz. In the spectral range here, we trace the $J$=5$\rightarrow$4 transition.

\begin{figure}
\begin{center}
\includegraphics[width=\textwidth]{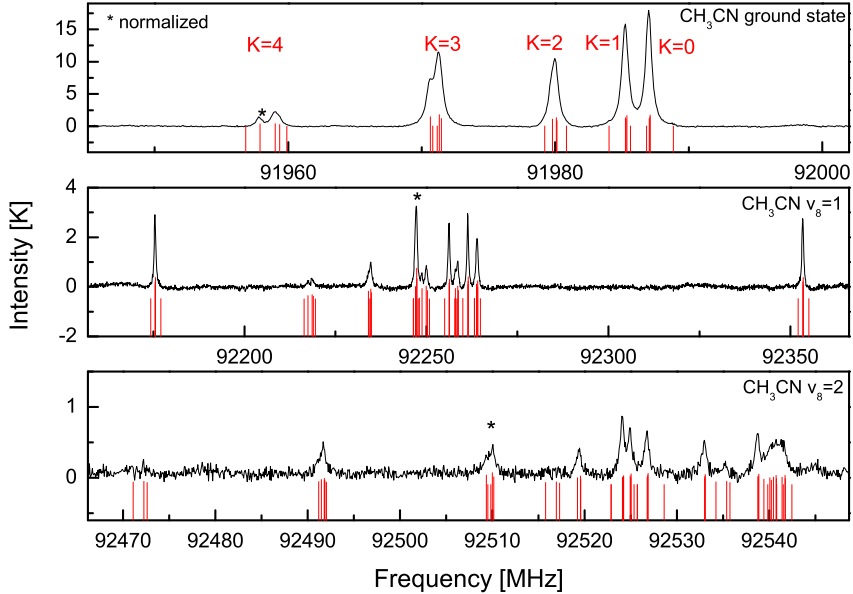}
\caption{\label{fig:emission} Emission spectrum of methyl cyanide.  Upper panel:  The emission spectrum of the ground vibrational state is clearly detected.  Middle panel:  The first vibrationally excited state of methyl cyanide is plotted ($v_8=1$).  Lower panel:  The second vibrationally excited state of methyl cyanide is detected ($v_8=2$).  In order to overplot the CDMS entries for each vibrational state, we normalized the CDMS intensities to the respective band marked in the spectra with an asterisk.  The total integration is 100 minutes and the pressure was kept constant at around 1~x~10$^{-2}$~mbar.  The RMS is in the order of 0.08~K.}
\end{center}
\end{figure}

Fig.~\ref{fig:emission} shows the emission spectrum of methyl cyanide derived through measurements of the molecular emission against a cold and a hot reference background. The resulting spectra are divided into three panels in Fig.~\ref{fig:emission}.
 In black the experimental spectrum is plotted, while the entries from the Cologne Database for Molecular spectroscopy (CDMS) are overplotted as red stick diagram (\cite{Endres:2016,Muller:2005,Muller:2001}).

 Each of the  three panels shows lines associated with rotational transitions of 
 CH$_3$CN in different vibrational levels.  The y-axes are calibrated using the above mentioned procedure and give the intensities in Kelvin. The x-axes show the frequency in MHz.  The rotational transitions within the ground vibrational state are shown in the upper panel, the rotational transitions within the vibrationally excited states, $v_8$=1 and v$_8$=2, are shown in the second and third panel, respectively.  For all vibrational states the $J$=5$\rightarrow$4 transitions are detected.  It can clearly be seen that quite a fraction of the population is distributed over the vibrational manifold even in such a small molecule as methyl cyanide.  Here, the first vibrationally excited state shows intensities in the order of one fourth of the intensities in the vibrational ground state.  Even the second vibrational excited state is clearly detected and again about one fourth of the intensities is seen in this state. The vibrational states differ by about 350~cm$^{-1}$ in energy each.

 CH$_3$CN shows a characteristic splitting of the different $K$-levels for each $J$ value ($|K|$~$<~J$). This $K$ splitting is obvious in the ground vibrational state and labelled in the first panel.  For some lines in the ground vibrational state a hyperfine splitting due to the nuclear spin of the nitrogen atom ($I$=1), is also resolved.  This can be seen best for the peak at 91960~MHz, which is marked with an asterisk.  Also, for the $K$~=~3 transition, around 91970~MHz, a shoulder caused by the splitting of the individual hyperfine components can be detected.  Here, a separation between the $F$=$J$ and $F$=$J$~$\pm$~1 components at 91970.7 and 91971.4~MHz can be seen.

 Apart from these spectroscopic details Fig.\ref{fig:emission} shows that substantial parts of the spectra can be recorded simultaneously with the emission spectrometer on an hour time scale. Much weaker signals are detectable thanks to the good S/N ratio of the measured spectra. In fact, the RMS of the intensities are in the order of 0.08~K and hence the spectrometer sensitivity meets our expectation (RMS=0.06~K, according to system calibration). This RMS results in a S/N of about 210 for the strongest transition within the ground vibrational state (first panel), a S/N of 38 for the strongest transition within the first vibrationally excited state (second panel), and a S/N of around 10 for the second vibrationally excited state (third panel).


  Each channel spacing of the XFFTS is around 76~kHz (2500~MHz divided by 32768 channels), while the spectral resolution is in the order of 88~kHz (equivalent noise bandwidth).  While this is sufficient to compare spectra to observations, it is not sufficient to fully resolve the hyperfine structure of (complex) molecules. The hyperfine components are partially resolved in our spectrum, as can be seen by the red stick diagram.  There are several reasons why we do not fully resolve the hyperfine components:  First:  Only if the spacing between individual hyperfine components is in the order of the spectral resolution of the XFFT-spectrometer this is possible.  This is not always the case for methyl cyanide as can be seen by the red stick diagram.  Second:  The Doppler broadening for 300~K warm lines is in the order of 200~kHz, and hence larger than the spectral resolution of the XFFTS.  Third:  Pressure broadening is not negligible in this spectrum and broadens the lines in addition to the Doppler broadening.

\begin{figure}[b]
\begin{center}
 \includegraphics[width=\textwidth]{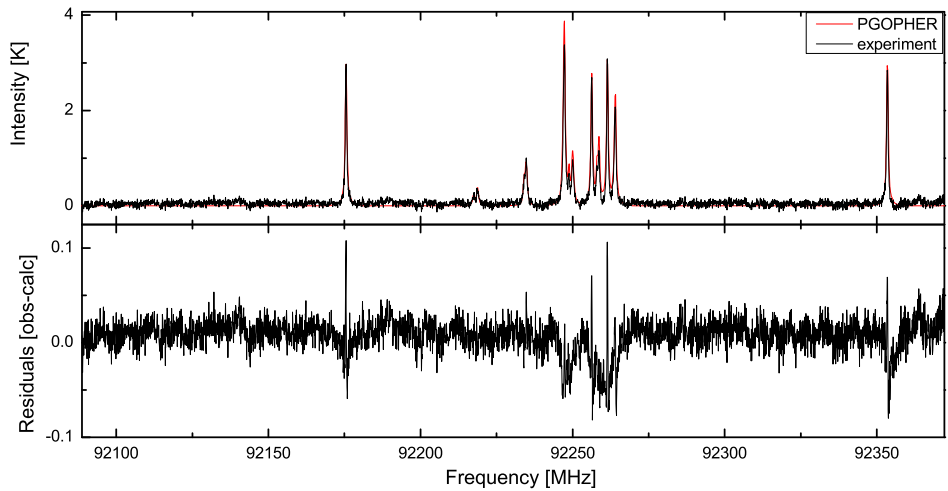}
 \caption{Emission spectrum of methyl cyanide.  Upper panel:  The emission spectrum of the $v_8=1$ is presented.  The experimental spectrum is plotted in black.  Overplotted is a PGOPHER simulation in red.  Lower panel:  The redisuals are shown when the simulation is subtracted from the experimental spectrum (obs-calc).  The simulation has been obtained using PGOPHER under LTE conditions and for 300~K.}
   \label{residuals}
\end{center}
\end{figure}

In order to show the frequency accuracy as well as the intensity accuracy obtained with our spectrometer, we simulate the expected spectrum under LTE (local thermodynamic equilibrium) conditions.  In Fig.~\ref{residuals} we plot the $v_8$=1 experimental spectrum (in black) together with a simulated spectrum (in red).  The simulation has been obtained using PGOPHER Version 10.0 (\cite{Western:2017}).  The program simulates the rotational structure by assuming LTE conditions.  Here, we used the molecular parameter files as given from the website of the Jet Propulsion Laboratory (JPL, \cite{Pickett:1998}).  The parameter file was downloaded for the first vibrational excited state and includes the hyperfine splitting for the nitrogen nucleus ($I$~=~1).  The temperature in the simulation is set to 300~K.  The only free parameter that was chosen in the simulation is the linewidth for a Voigt line profile.

We assume Doppler broadening for our lines, which can be calculated to be in the order of FWHM=200~kHz at a center frequency of 92~GHz, applying the following formula:

\begin{equation}
\rm{\Delta \nu = 2 \sqrt{ln(2)} \times \nu_0~c^{-1}~\sqrt{\frac{2 kT}{m}}}
\end{equation}

Here, $\nu_0$ is the center frequency (here around 92~GHz), c is the speed of light, k indicates the Boltzmann constant, T is around room-temperature (300~K) and m is the molecular mass.  Hence, we chose for a Gaussian component of 0.2~MHz in the simulation and subsequently added the Lorentzian lineshape to the simulation.  The best results were thus obtained using 0.2~MHz Gaussian linewidths, in addition to 0.6~MHz Lorentzian linewidths.

The result can be seen in Fig.~ \ref{residuals}.   The intensities are calibrated and given in Kelvin.  In the simulation (red) the intensities are normalized to the intensity obtained in the emission experiment. Hence, for normalization purposes the emission line at around 92260 MHz was chosen. The frequency scale is given in MHz.   The lower panel shows the resulting residuals when subtracting the simulated PGOPHER spectrum from the experimental spectrum (obs - calc).  The intensity axis is still in Kelvin.  Here, the residuals are in the order of a few percent (about 2.5~$\%$) for the strongest lines.  For weak emission lines (92220--92240~MHz), the residuals cannot be discriminated from the baseline noise.

  \section{Conclusion}

  In this paper we demonstrate the capabilities and limitations of a home-built 100~GHz
room-temperature heterodyne receiver.

  \begin{itemize}
  \item  The Allan stability was determined to be in the order of 100~s for 76~kHz channel width.
  \item  The receiver noise temperature (at room-temperature) is around 400~K, and meets the expectations. 
   \item The system noise temperature is in the order of 600~K and should be improved through better optics arrangement.
  \item  Emission spectra of methyl cyanide show the  $J$=5$\rightarrow$ 4 rotational transitions in the ground vibrational state, as well as the first and second excited vibrational state $v_8$=1 and $v_8$=2, respectively.
  \item  The spectral accuracy, i.e., the frequency and intensity accuracy was evaluated by comparing the methyl cyanide spectra obtained from the emission experiment to simulations obtained from PGOPHER.  The residuals of the intensities were in the order of a few percent for the strongest deviation from the simulation.
\end{itemize}

 We conclude that the new technology that can now be used in the laboratory environment has advantages with respect to the traditional frequency modulation (FM) absorption spectroscopy.  Given the fact that broadband spectrometers up to 16~GHz are currently available, this will result in even more efficient ways of obtaining spectra of complex organic species over large frequency ranges. 
  Relative intensity measurements show a reproducibility of the emission spectra much better than the few percent reported in our comparison between experiment and theoretical predictions. Thanks to this high quality intensity information it is foreseen that temperature dependent intensity measurements can be employed to derive molecular energy diagrams without associated rotational model Hamiltonians as in the traditional bootstrap approach (\cite{Medvedev:2007,DeLucia:2010}). This is thought to open the door to increased analysis speeds with many more molecular levels being included. This development is urgently needed in the era of ALMA where very complex spectra with unprecedented sensitivity and thus many molecular levels need to be understood.
  Since such species show a spectrum in a wide frequency regime with thousands of lines, it would be preferable to use more efficient detection methods as currently possible with FM technique.  Coupling of laboratory emission receivers with other techniques, such as chirped-pulse excitation, would also result in an increased S/N ratio. First experiments along these lines demonstrate increased sensitivity with detection of even higher vibrational states of methyl-cyanide.

Currently, tests are underway to cool-down the low-noise amplifier to cryogenic temperatures, which also will decrease the system noise temperature and hence increase the spectrum acquisition efficiency.  Furthermore, a new spectrometer making use of SIS-receivers (superconductor-insulator-superconductor) at 300~GHz is tested in our laboratory, which shows promising application for the acquisition of spectra of complex organic molecules with astrophysical importance.

To this date the fast response time of the emission detection has not been used except in first chirped-pulse experiments. However, these capabilities may also be used to identify transient species such as radicals and ions which are a challenge to laboratory work and astrophysics as well. 
All together the emission spectrometers developed in several laboratories around the globe will allow new experiments including the study of the influence of collisions or energy transfer on molecular spectra or the detection of molecules in non-standard environments such as desorbing molecular ices or products of chemical reactions.

\begin{acknowledgments}
The emission spectrometer development is carried out within the Collaborative Research Centre 956, "Conditions and Impact of Star Formation", sub-projects B3, B4 and S, funded by the Deutsche Forschungsgemeinschaft (DFG). M.H. acknowledges funding through the Bonn-Cologne-Graduate-School for physics and astronomy (BCGS).
We would like to thank Gundolf Wieching at MPIfR Bonn for fruitful discussions.
\end{acknowledgments}

\end{document}